\documentclass[journal,twoside,web]{ieeecolor}
\usepackage{tmi}
\usepackage{cite}
\usepackage{amsmath,amssymb,amsfonts}
\usepackage{algorithmic}
\usepackage{graphicx}
\usepackage{textcomp}
\usepackage{multirow}
\usepackage{amssymb}
\usepackage{bbm}
\usepackage[misc]{ifsym}
\usepackage{graphicx}
\usepackage{multirow}
\usepackage{float}
\usepackage{ulem}
\usepackage{amsmath}
\usepackage{color}
\usepackage{hyperref}
\usepackage{multirow}
\usepackage{booktabs}
\usepackage{makecell}

\def\BibTeX{{\rm B\kern-.05em{\sc i\kern-.025em b}\kern-.08em
    T\kern-.1667em\lower.7ex\hbox{E}\kern-.125emX}}
\begin{document}
\title{Enhanced MRI Representation via Cross-series Masking}

\author{Churan Wang, Fei Gao, Lijun Yan, Siwen Wang, Yizhou Yu
, Yizhou Wang}

\maketitle

\begin{abstract}
Magnetic resonance imaging (MRI) is indispensable for diagnosing and planning treatment in various medical conditions due to its ability to produce multi-series images that reveal different tissue characteristics. However, integrating these diverse series to form a coherent analysis presents significant challenges, such as differing spatial resolutions and contrast patterns meanwhile requiring extensive annotated data, which is scarce in clinical practice. Due to these issues, we introduce a novel Cross-Series Masking (CSM) Strategy for effectively learning MRI representation in a self-supervised manner. Specifically, CSM commences by randomly sampling a subset of regions and series, which are then strategically masked. In the training process, the cross-series representation is learned by utilizing the unmasked data to reconstruct the masked portions. 
This process not only integrates information across different series but also facilitates the ability to model both intra-series and inter-series correlations and complementarities. With the learned representation, the downstream tasks like segmentation and classification are also enhanced. Taking brain tissue segmentation, breast tumor benign/malignant classification, and prostate cancer diagnosis as examples, our method achieves state-of-the-art performance on both public and in-house datasets.

\end{abstract}


\section{Introduction}
\label{sec:introduction}
\begin{figure}[t!]
\begin{center}
\includegraphics[width=0.49\textwidth]{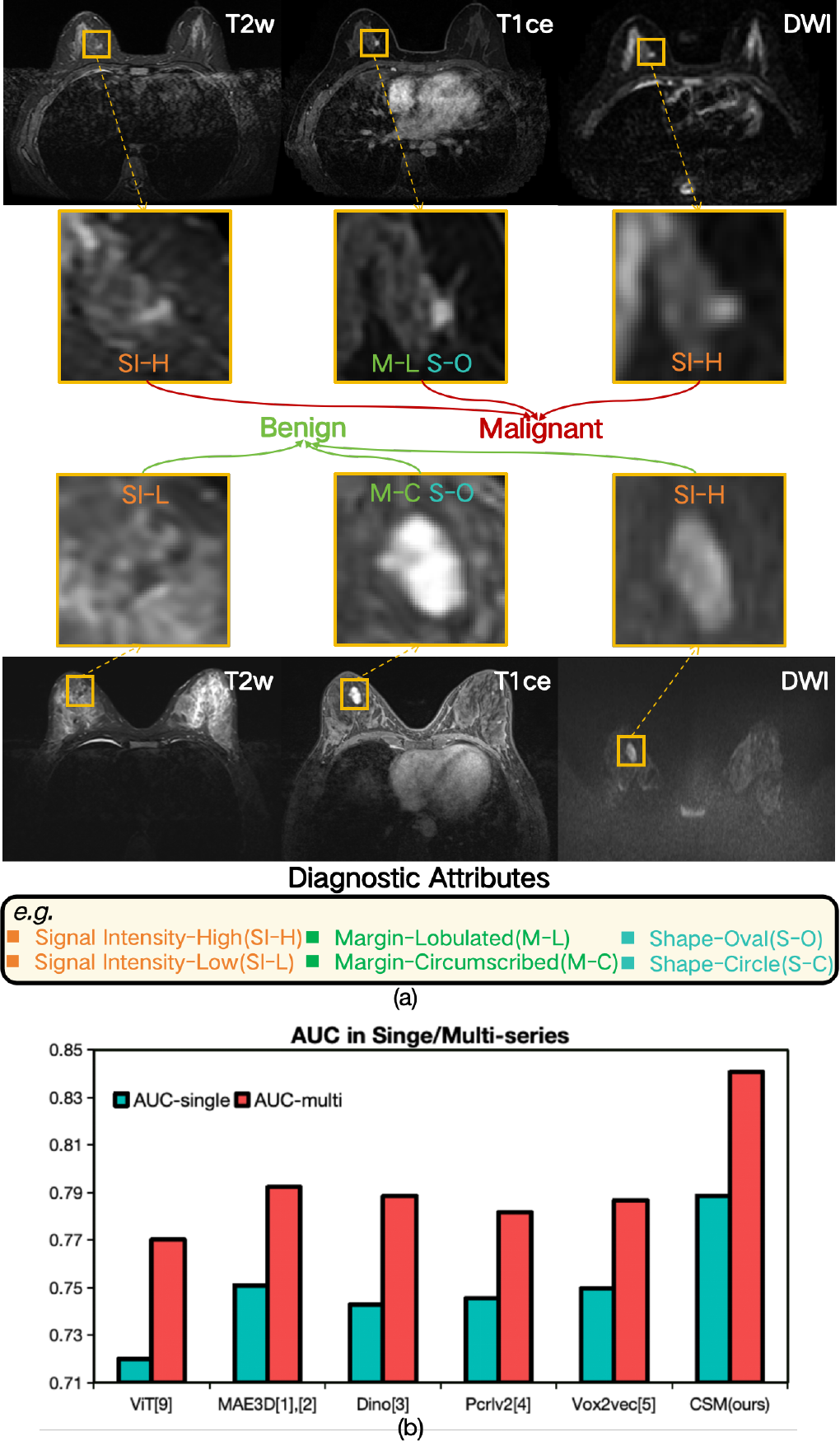}
\end{center}
\centering
\caption{Examples of using multiple series for clinical diagnosis, take breast malignant diagnosis as an example. (a) The upper case is malignant (high intensity in T2w and DWI, oval shape and lobulated margin in T1ce) and the lower case is benign (low intensity in T2w, high intensity DWI, oval shape and circumscribed margin in T1ce), demonstrating the three series together with a close-up view of the lesion and its surrounding area. Only considering diagnostic attributes in a single series can not diagnose lesions. (b) The effectiveness of using multiple series is much higher than that of using a single series. Our method is state-of-the-art.
}
\label{motivation}
\end{figure}

Magnetic resonance imaging (MRI) is a widely used medical technology that can provide rich and diverse information about the anatomy and pathology of human tissues \cite{mriback4,bookmri}. 
MRI can produce different types of images, called \textit{series}, by varying many factors such as the acquisition parameters and the contrast mechanisms \cite{mriback4,mriback5,mriback6_prostate}. As shown in Figure~\ref{motivation}(a), different series can highlight different aspects of the same tissue, such as morphology, function, metabolism, or diffusion. And Figure~\ref{motivation}(b) gives the quantitative comparisons and illustrates that combining multiple series can achieve higher performances. Therefore, multi-series MRI analysis is essential for accurate and comprehensive diagnosis \cite{mriback1,mriback2,mriback3}. 


However, multi-series MRI analysis poses significant challenges in medical image computing. One of the main challenge of multi-series representation learning is \textbf{how to effectively integrate and fuse the information from different series}, which may have different factors such as spatial resolutions, signal-to-noise ratios, intensity distributions, and contrast patterns \cite{mriback4}. Traditional methods often rely on hand-crafted features or predefined rules to combine different series, which may not capture the complex and subtle relationships among them \cite{old1,old2,old3}. Deep learning methods have shown great potential in learning high-level and abstract features from multi-series data. For example, Zhang \textit{et al.} \cite{tmiattention} propose a deformable aggregation module for multiseries feature fusion. A2FSeg \cite{miccaiattention} design two stages of feature fusion, including a simple average fusion and an adaptive fusion based on an attention mechanism. However, these methods often require a substantial amount of annotated data. The acquisition of such annotated data is frequently challenging due to the demands for time, cost, and specialized expertise. Compared to annotated data, unannotated data is more readily accessible in clinical practice.

To fully leverage unlabeled data to enhance model performance, self-supervised learning (SSL) has emerged as a promising approach in recent years. Initially, SSL has demonstrated significant performances in natural images~\cite{hard_wang,mixedae,simple_chen,momentum_he,bootstrap_grill,siamese_chen}. In terms of supervision types, the current methods are primarily divided into two major categories: contrast-based and reconstruction-based approaches. The core idea of Contrast-based methods such as DINO~\cite{dino} is to minimize the feature distances between different views of the same image while maximizing the distances between features of different images, thereby forcing the model to learn discriminative representations for different instances. However, contrast-based methods primarily focus on learning global representations but lack in capturing local features under single series. This limitation results in suboptimal performance on dense prediction tasks such as segmentation and lack of multi-series representation learning. 
Reconstruction-based methods are primarily centered around mask image modeling~\cite{wang2024fremim}. By masking a significant portion of the image content for reconstruction, the model is capable of learning both comprehensive global and fine-grained local representations. However, existing approaches such as MAE3D~\cite{mae} and the work of Zhou \emph{et al.}~\cite{zhouisbi} have only concentrated on single image series representation, but ignoring the potential of multi-series representation.


Due to the scarcity of annotated data, SSL has also garnered significant attention in the field of medical image analysis~\cite{modelgenesis,rubikscube,modelgenesis,swinunetr}. Initially, researchers in this field have also designed proxy tasks~\cite{xie2020pgl,ye2022desd,zhou2021preservational} that focus on the characteristics of medical images, such as Vox2vec~\cite{vox2vec} and Pcrlv2~\cite{pcrlv2}. The core strategy they use is to apply various perturbations to the original images and then perform precise restoration, which enables the model to learn fine-grained image features. Despite being tailored for medical imaging, existing methods are limited to single series, neglecting the crucial aspect of multi-series self-supervised representation learning. This oversight limits the potential for models to capture comprehensive features inherent in the diverse information present across multiple series.
To address this, we propose a SSL method to enhance multi-series representation learning by \textbf{C}ross-\textbf{S}eries \textbf{M}asking (\textbf{CSM}) including intra-series and inter-series masking strategies. Specifically, for intra-series masking, each series is independently masked of a portion of regions. By utilizing the mask-image-modeling strategy with multiple series that have been randomly masked, the target is to reconstruct the masked content within each series. This process enables the model to learn the commonalities and complementarities between different series, as well as the contextual information within each individual series. Furthermore, for inter-series masking, some series are randomly and entirely masked, and these series are reconstructed based on the other series. This allows the model to learn the distinctive features of each series as well as the inter-series relationships. By combining these two masking strategies, the model can acquire powerful multi-series representations that can benefit downstream tasks. 
Compared with barely using a single series, employing multi-series images can enhance the models' diagnostic performance in our proposed method as well as in other methods. Additionally, our method outperforms other compared baselines, which is displayed in the bar chart in Figure~\ref{motivation}(b).

Our main contributions are as follows:

- We propose a novel self-supervised learning method for multi-series MRI representation learning, which uses cross-series masking as a preset task to learn the representation that can fuse the information from different series.

- We meticulously design a cross-series masking strategy that encompasses both intra-series and inter-series masking approaches. By employing mask image modeling to reconstruct the original images, the model can learn comprehensive multi-series representations.

- We conduct extensive experiments on two public datasets and one in-house dataset. Our method can achieve state-of-the-art results on various downstream tasks, such as brain tissue segmentation, breast tumor malignant diagnosis, and prostate tumor malignant diagnosis. We also show that our method can effectively reduce the annotation cost.

\section{Related Work}
\subsection{Self-supervised learning (SSL)}
The core idea of self-supervised learning (SSL) is to leverage the inherent characteristics of the data to obtain data representations, rather than relying on manually labeled data. In recent years, prevalent self-supervised learning (SSL) methods roughly reveal two dominant categories: contrast-based and reconstruction-based. 

\noindent\textbf{Contrast-based} Contrast-based methods focus on learning data features from the contrastive information of data samples\cite{dino,vox2vec,simple_chen,big_chen,empirical_chen,bootstrap_grill,momentum_he,siamese_chen,unsupervised_caron,selfsupervised_wen} by contrastive learning which is one of the common methods in SSL
.
For example, Dino \cite{dino} introduced the self-distillation strategy and avoided model collapse by the centering and sharpening operations. Pcrlv2 \cite{pcrlv2} proposed by Zhou \emph{et al.} conducted multi-scale feature comparison to learn multi-scale representations.
This demonstrated that contrastive learning could achieve significant results without explicit negative samples.


With the the success of Transformers in natural language processing, Transformers have also begun to explore application to visual tasks in SSL. Chen \emph{et al.}\cite{empirical_chen} conducted empirical studies on self-supervised training of vision transformers, providing valuable insights into the behavior of these models in a self-supervised setting.

In addition, Goncharov \emph{et al.}\cite{vox2vec} proposed the Vox2Vec framework, designed to capture fine-grained local features in medical images, thereby learning rich voxel-level representations. By designing a specific contrastive loss function, this framework trains the model to bring different modalities of images from the same patient or different views of the same image closer in the feature space, while pushing features from different patients or different images further apart.


Overall, these works illustrate the diversity and richness of contrastive learning in self-supervised visual representation learning. However, the predominant focus on single-image series presents a limitation. This suggest an promising frontier in harnessing the synergistic potential of multi-series data for richer, more nuanced representations.

\noindent\textbf{Reconstruction-based}
Reconstruction-based methods employ a masked image modeling strategy, wherein portions of the input image are obscured, prompting the model to predict the missing parts. This strategy compels the model to delve into the intrinsic structure of the image~\cite{zhouisbi,mae,pcrlv2,simple_xie,hard_wang,peco_dong,neural_oord,masked_wei,semmae_li,adversarial_shi}, thereby leading to robust and comprehensive representations.


Oord \emph{et al.}\cite{neural_oord} introduced neural discrete representation learning, which laid the theoretical foundation for masked image modeling. He \emph{et al.}~\cite{mae} introduced a framework which learns feature representations through a regression task that predicts the original pixel values. 
Similarly, Zhou \emph{et al.}\cite{zhouisbi} proposed a scalable encoder that learns representations by randomly masking patches of input images and reconstructing the missing pixels, effectively extending the concept that He \emph{et al.}~\cite{mae} proposed to 3D medical image.


However, the aforementioned SSL methods generally share a common drawback: they are designed for single images and do not consider the issue of information fusion across multiple series. MRI scans typically consist of multiple series, each emphasizing different types of information. To make accurate judgments, it is crucial to integrate information from various series, which requires a model capable of handling multi-series data.
Our proposed model overcomes this limitation by flexibly accommodating different numbers and types of series. It achieves superior performance across multiple datasets and various downstream tasks, demonstrating its ability to effectively process and integrate cross-series information.


\subsection{Multi-series representation learning}
In clinical practice, multi-series MRI plays a crucial role in diagnosing various diseases. Common MRI series include T1-weighted imaging, T2w-weighted imaging, dynamic contrast-enhanced T1-weighted imaging, and diffusion-weighted imaging (DWI). Studies have underscored the significance of this approach in diagnosing a spectrum of conditions, with notable research advancements in areas such as prostate cancer, bladder cancer, breast cancer, and glioblastoma, as documented in the literature~\cite{prostate_Roussel, prostate_Wysock, bladder_Panebianco, breast_mann, breast_pinker, glioblastoma_Shukla}. Furthermore, the research in~\cite{mpmri_Winfield, mpmri_Omari} has illuminated the broader applicability of multi-series MRI, highlighting its potential to augment diagnostic accuracy across various medical disciplines. The integration of data from multiple MRI series has consistently demonstrated enhanced diagnostic efficacy, a finding corroborated by the prevalence of these diseases on a global scale~\cite{global_cancer2022}. These insights underscore the broader relevance of multi-series representation learning in advancing medical imaging diagnostics.

For example, Roussel \emph{et al.}\cite{prostate_Roussel} have mentioned that contemporary multi-series MRI protocols provide anatomical insights and offer qualitative, semi-quantitative, and fully quantitative imaging biomarkers that correlate with histological subtypes, tumor grades, and clinical behavior. Winfield \emph{et al.}\cite{mpmri_Winfield} details the essential role of multi-series MRI in three high-incidence but diagnostically challenging diseases: prostate cancer, breast cancer, and glioblastoma. The sensitivity or specificity of single-series MRI for these diseases is relatively low (e.g., the specificity for breast cancer is at most around 71\%), and clinical practice tends to favor multi-series MRI to enhance diagnostic performance.

In summary, multi-series MRI significantly improves diagnostic accuracy by integrating diverse imaging information, making it an indispensable tool in clinical diagnostics. Compared to single-series MRI, the importance of combining multiple series lies in its ability to provide more comprehensive and precise diagnostic information, thus playing a crucial role in the early detection and accurate diagnosis of diseases.

\begin{figure*}[t!]
\begin{center}
\includegraphics[width=\textwidth]{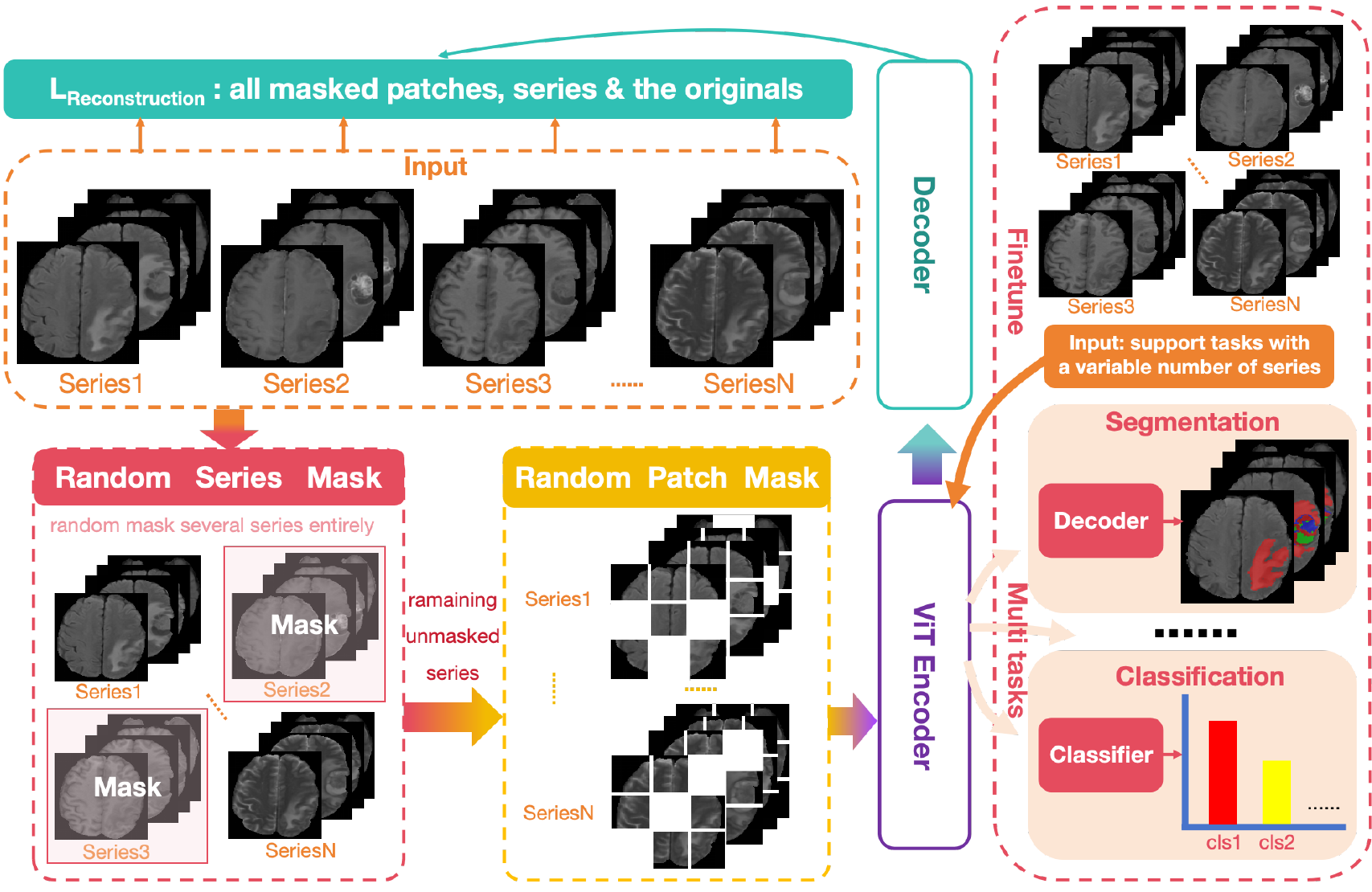}
\end{center}
\centering
\caption{The overview of our proposed method for learning MRI representation via cross-series masking. It begins with multiple MRI series inputs, which undergo random series and patch masking. The remaining unmasked regions are then encoded by a ViT Encoder and learned to reconstruct the masked series and patches. The well-trained encoder can model the correlations between multiple series and extract the cross-series representation. Based on this, the performance of other downstream tasks can be enhanced without increasing the number of training data.}
\label{method_overview}
\end{figure*}

\section{Method}


Multi-series fusion plays an important role in MRI analysis. Each MRI series can provide unique and complementary information about the tissues. However, these series differ in many aspects, such as acquisition parameters, spatial resolutions, and contrast mechanisms. Traditional supervised learning methods rely heavily on labeled data. To address these issues, we propose a self-supervised learning strategy that employs cross-series masking to learn the representation from multi-series images. We can efficiently fine-tune the encoder with a task-specific head for each downstream task using even a small set of annotated data. 

\subsection{Problem Setup.} 
Our dataset contains $\{x^i,y_i\}_{i \in \{1,...,n\}}$, in which $X^i$ denotes the $i$th patient that has $s$ series images $x^i ({x^i}_1, {x^i}_2,...,{x^i}_s)$, $y_i$ denotes the ground truth in different downstream tasks (\emph{e.g.}, lesion regions from manual annotations in segmentation tasks, the binary disease label in malignant diagnosis tasks). 

Figure~\ref{method_overview} outlines the overall pipeline of our method. As shown, our method contains two stages: \textit{\textbf{Self-supervised Representation Learning (SSL)}} stage and \textit{\textbf{Downstream Fine-tuning}} stage. During the \textit{\textbf{SSL}} stage. In the \textit{\textbf{SSL}} stage, only the image data is provided for general image representation learning without any annotation. During the \textit{\textbf{Downstream Fine-tuning}} stage, we utilize the pre-trained encoder in \textit{\textbf{SSL}} stage to fine-tune specific downstream tasks using manually annotated data. This process allows us to achieve great results even with a minimal amount of data. 

\begin{figure}[ht]
\begin{center}
\includegraphics[width=0.48\textwidth]{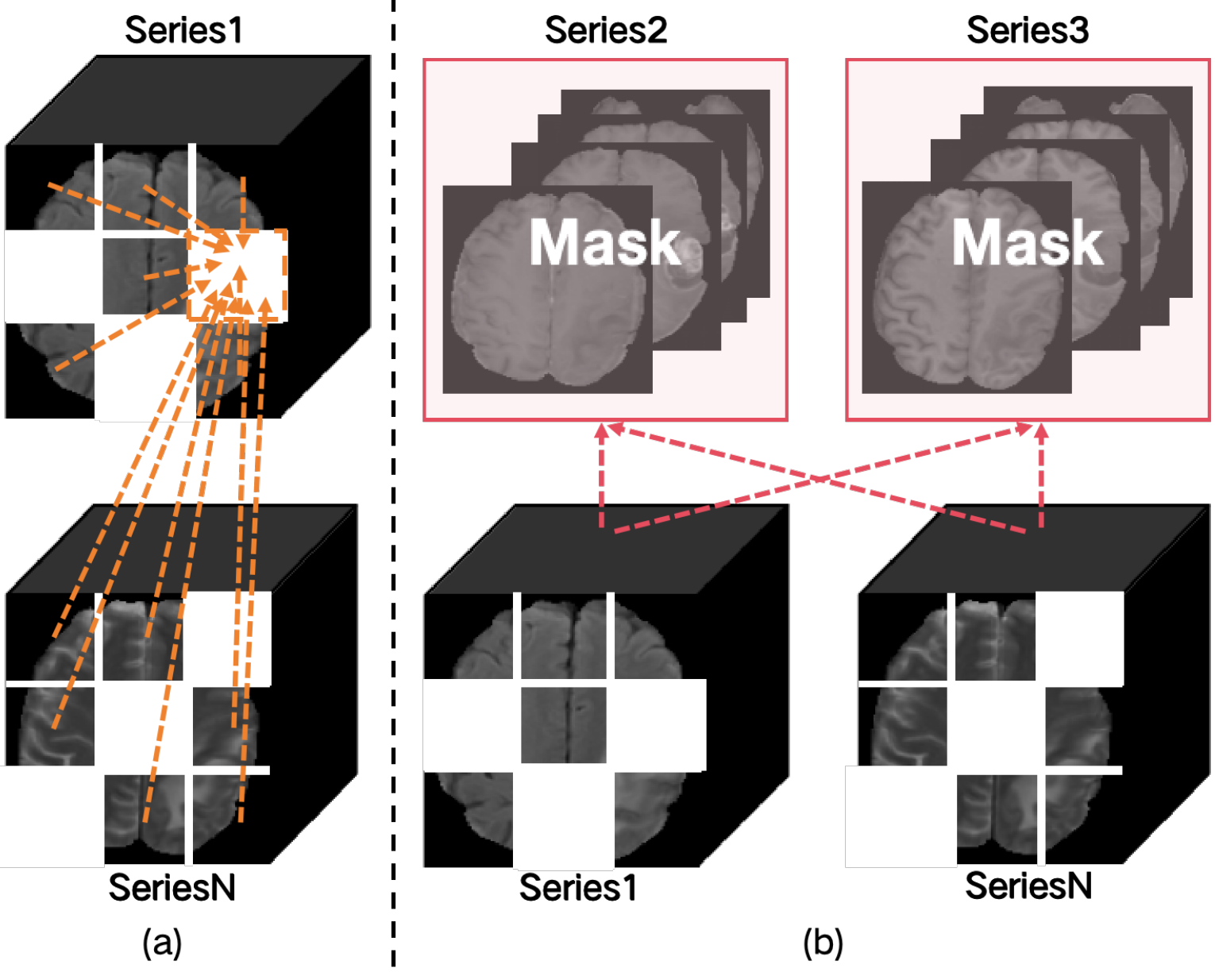}
\end{center}
\centering
\caption{The strategy of our proposed CSM method. 
(a) Intra-series masking, which randomly masks a substantial proportion of the patches within each series respectively. Adopting the intra-series masking strategy, the model can not only leverage the context from both inter-series and intra-series but also learn the inter-series complementarities. 
(b) Inter-series masking, which randomly fully masks a subset of the series. This strategy forces the model to learn the complementarity and relationship in a global view.}
\label{mask_strategy}
\end{figure}

\subsection{Self-supervised Representation Learning.}
In this stage, our goal is to learn the representation that captures the essential features across multiple series. Thus we design a \textit{\textbf{cross-series masking strategy}} including intra-series masking and inter-series masking, shown in Figure~\ref{mask_strategy}. Each series is divided into N non-overlapping patches, with each patch representing a visual token. 

For intra-series masking, we randomly mask a substantial proportion of the tokens within each series, which is depicted in Figure~\ref{mask_strategy}(a). The random masked patches are denoted as set $P_m$, and the remaining unmasked patches are denoted as set $P_u$. In this strategy, each series is subjected to independent random masking, meaning that the regions masked in each series are distinct from one another. The motivation behind this design is that, for a given invisible token, its reconstruction primarily relies on other regions within the current series, and the corresponding regions in other series. In this way, the model can first learn the context within the same series, and on the other hand, it can learn the interdependence and complementarities between different series. However, if all series are masked in the same regions, the model can only learn the intra-series and inter-series context. This limitation would prevent the model from grasping the inter-series relationships and complementarities which are essential for effective multi-series representation learning. Consequently, such a deficiency would impede the model's performance in capturing the comprehensive information present across different MRI series.

Regarding inter-series masking, we randomly select a subset of the series and fully mask them, meaning that the mask ratio for these series is 100\%, which is outlined in Figure~\ref{mask_strategy}(b). The number of masked series is $k$, which is a random variable ranging from $[1, s-1]$. The masked series are denoted as set $S_m$. By reconstructing these fully masked series, it is akin to forcing the model to address the issue of series absence, thereby learning the overall commonalities and complementarities between different series. The aforementioned intra-series masking primarily focuses on learning local commonalities and complementarities, whereas inter-series masking is holistic, posing a higher level of difficulty and compelling the model to learn stronger inter-series representations. 

By combining these two masking strategies, the model can learn comprehensive features within and between series from both local and global perspectives.


Specifically, we employ an encoder $E_{\theta_{\mathrm{enc}}}$, which processes the unmasked patches $P_u$, to produce a latent representation. A decoder $D_{\theta_{\mathrm{dec}}}$ then attempts to reconstruct the masked patches $P_m$ and series $S_m$ from this representation: 
\begin{align}
    \label{eq:forward}
    \hat{x}=D_{\theta_{\mathrm{dec}}}(E_{\theta_{\mathrm{enc}}}(P_u))
\end{align}

The reconstruction loss $\mathcal{L}_{\mathrm{reconstruction}}$ is defined to measure the differences between the reconstructed series and the original unmasked series in mean squared error:

\begin{align}
    \label{eq:rec}
    \mathcal{L}_{\mathrm{reconstruction}}(\theta_{\mathrm{enc}},\theta_{\mathrm{dec}}) := \sum_{x \in \mathrm{clus}(P_m, S_m)} \Vert x - \hat{x}(\theta_{\mathrm{enc}},\theta_{\mathrm{dec}}) \Vert_2 
\end{align}

By minimizing this loss, the encoder-decoder pair learns to recover the complete set of series from partially observed data. Through this self-supervised learning stage, our CSM acquires the capability to infer missing information and generate a comprehensive representation of multi-series MRI data, which is crucial for downstream tasks.

\subsection{Downstream Fine-tuning.}
After the above stage, we obtain a pre-trained encoder $f_{\theta_{\mathrm{enc}}}$ that has learned a rich representation of the multi-series MRI images. The next step is to adapt these representations to specific downstream tasks, such as segmentation and classification, which we denote as $T$.

Given a small set of annotated data $D_{T}=\left\{\left(x_{i}, y_{i}\right)\right\}_{i=1}^{N}$, we fine-tune the encoder along with a task-specific head $H_T$ by solving the following problem: 
\begin{align}
    \label{eq:downstream}
    \min _{E, H_{T}} \frac{1}{N} \sum_{i=1}^{N} L_{T}\left(H_{T}\left(E\left(x_{i}\right)\right), y_{i}\right)
\end{align}
$L_T$ is the loss function relevant to the task $T$, such as cross-entropy for classification or Dice coefficient for segmentation. The task-specific head $H_T$ can be a variable set of layers that map the features extracted by the encoder $E$ (initialized by $E_{\theta_{\mathrm{enc}}}$) to the output space of the task $T$. 

Through this adaptation process, the encoder fine-tuned on task-specific annotated data is expected to achieve superior performance on the target task, demonstrating the practical utility of the learned representations.

\section{Experiments}

\subsection{Datasets \& Implementation}

\begin{figure}[t!]
\includegraphics[width=0.5\textwidth]{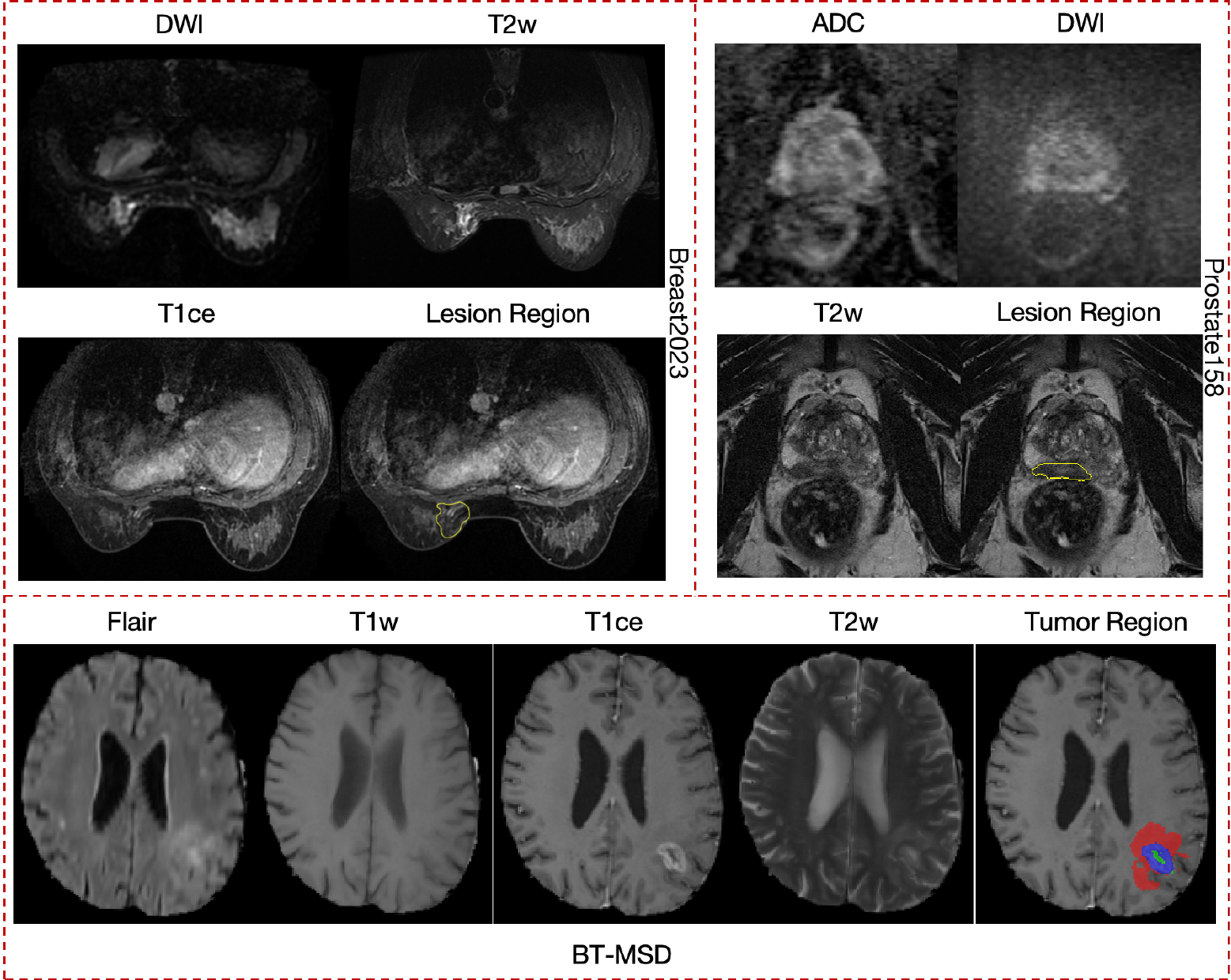}
\centering
\small
\caption{Examples from the datasets we use in our study. \textbf{Breast2023} is our in-house breast MRI dataset, which contains T1ce, DWI and T2w series for each patient. The lesion is delineated by yellow lines, and its malignancy is confirmed by pathological test. \textbf{BT-MSD} is a publicly available brain MRI tumor segmentation dataset with four modalities(T1w, T1ce, T2w, Flair), the tumor region area is delineated by experts. \textbf{Prostate158} is a public dataset about prostate MRI. Each patient includes three series - T2w, DWI, and ADC, with the cancerous area marked by the doctor (the part within the yellow curve).}
\label{dataset_example}
\end{figure}

To evaluate the effectiveness of our method CSM, we verify it on brain tissue segmentation, breast tumor benign/malignant classification, and prostate cancer diagnosis (benign/malignant classification). We consider both the public dataset Brain Tumour task in Medical Segmentation Decathlon (BT-MSD) Challenge \cite{brats}, the in-house dataset Breast2023, and a public dataset Prostate158 as shown in Figure~\ref{dataset_example}. 

\noindent\textbf{BT-MSD.} BT-MSD has 484 multi-series MRI brain scans (Flair, T1w, T1ce, and T2w)\cite{zhouisbi,brats}, which provides comprehensive information about brain structures and tumoral information. The dataset includes manual segmentations by radiologists, identifying enhancing tumor regions (ET), necrotic tumor core regions (NCR), and peritumoral edematous/invaded tissues (ED). During the evaluation process, we use the Dice coefficients for three categories: tumor core (TC), whole tumor (WT), and enhancing tumor (ET), where tumor core = NCR, whole tumor = ET + NCR + ED, and enhancing tumor = ET.

\begin{figure}[t!]
\includegraphics[width=0.48\textwidth]{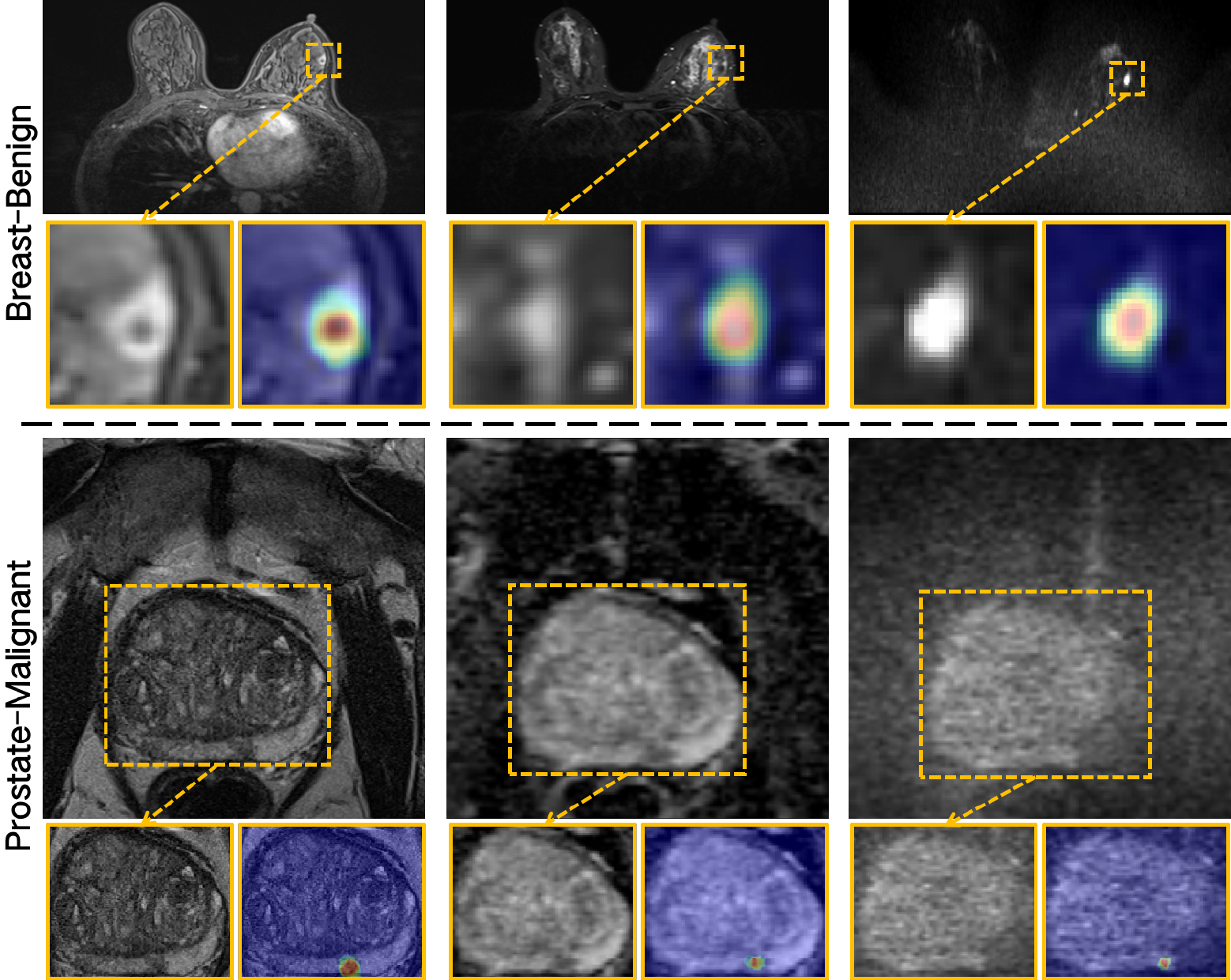}
\centering
\caption{Examples of class active maps of different series in Breast2023 and Prostate under multi-series classification.}
\label{cam}
\end{figure}

\noindent\textbf{Breast2023.} Breast2023 is an in-house dataset that includes multi-series MRI images from 515 lesions, which contains t1ce, DWI, and T2w series. Each lesion’s benignity or malignancy has been pathologically confirmed. Based on the delineation by radiologists, We extract the lesion areas from each MRI series. To leverage the edge information of the lesions, we slightly expand the lesion regions to include the surrounding areas of the lesions.

\begin{table*}[!t]
\caption{MSD Brain Tumor Segmentation using multi-series. WT denotes the whole tumor, ET denotes the enhancing tumor, and TC denotes the tumor core.}
\centering
\begin{tabular}{c|c|c|c|c}
\hline
 Methodology & Average & WT & ET & TC  \\
\hline
Unetr \cite{unetre} & $0.7805 \pm 0.022$ & $0.9137 \pm 0.021$ & $0.6246 \pm 0.025$ & $0.8031 \pm 0.020$ \\
MAE3D \cite{zhouisbi,mae} & $0.7904 \pm 0.015$ & $0.9156 \pm 0.014$ & $0.6172 \pm 0.018$ & $0.8383 \pm 0.012$ \\
Dino \cite{dino} & $0.7902 \pm 0.018$ & $0.9148 \pm 0.019$ & $0.6195 \pm 0.017$ & $0.8363 \pm 0.018$ \\
Pcrlv2 \cite{pcrlv2} & $0.7880 \pm 0.016$ & $0.9037 \pm 0.013$ & $0.6141 \pm 0.018$ & $0.8463 \pm 0.016$ \\
Vox2vec \cite{vox2vec} & $0.7877 \pm 0.015$ & $0.9053 \pm 0.015$ & $0.6103 \pm 0.014$ & $0.8478 \pm 0.015$ \\ \cline{1-5}
\textbf{CSM (Ours)} & $\textbf{0.8345} \pm 0.007$ &  $\textbf{0.9346} \pm 0.004$ & $\textbf{0.6738} \pm 0.009$ & $\textbf{0.8951} \pm 0.007$ \\ \cline{1-5}
Zhang \textit{et al.} \cite{tmiattention} & $0.7932 \pm 0.011$ &  $0.8985 \pm 0.009$ & $0.6736 \pm 0.010$ & $0.8074 \pm 0.013$ \\
A2FSeg \cite{miccaiattention} & $0.7941 \pm 0.010$ &  $0.9089 \pm 0.008$ & $0.6210 \pm 0.011$ & $0.8525 \pm 0.012$ \\\cline{1-5}
Zhang \textit{et al.} \cite{tmiattention} + \textbf{CSM (Ours)} & $\textbf{0.8351} \pm 0.005$ &  $\textbf{0.9342} \pm 0.004$ & $\textbf{0.6755} \pm 0.006$ & $\textbf{0.8955} \pm 0.006$ \\
A2FSeg \cite{miccaiattention} + \textbf{CSM (Ours)}& $\textbf{0.8462} \pm 0.005$ &  $\textbf{0.9351} \pm 0.005$ & $\textbf{0.7012} \pm 0.006$ & $\textbf{0.9024} \pm 0.005$\\

\hline
\end{tabular}

\label{tab:MSD-multi}
\end{table*}

\begin{table*}[!t]
\caption{MSD Brain Tumor Segmentation using single series. WT denotes the whole tumor, ET denotes the enhancing tumor, and TC denotes the tumor core.}
\centering
\begin{tabular}{c|c|c|c|c}
\hline
 Methodology & Average & WT & ET & TC  \\
\hline
Unetr \cite{unetre} & $0.7667 \pm 0.025$ & $0.8970 \pm 0.028$ & $0.6022 \pm 0.021$ & $0.8009 \pm 0.025$\\
MAE3D \cite{zhouisbi,mae} & $0.7832 \pm 0.019$ & $0.9077 \pm 0.016$ & $0.6289 \pm 0.019$ & $0.8131 \pm 0.021$ \\
Dino \cite{dino} & $0.7783 \pm 0.021$ & $0.9070 \pm 0.020$ & $0.6021 \pm 0.019$ & $0.8259 \pm 0.023$ \\
Pcrlv2 \cite{pcrlv2} & $0.7667 \pm 0.016$ & $0.8804 \pm 0.019$ & $0.5975 \pm 0.014$ & $0.8220 \pm 0.015$ \\
Vox2vec \cite{vox2vec} & $0.7739 \pm 0.017$ & $0.8970 \pm 0.017$ & $0.5941 \pm 0.019$ & $0.8305 \pm 0.014$ \\ \cline{1-5}
\textbf{CSM (Ours)} & $\textbf{0.8013} \pm 0.008$ &  $\textbf{0.9089} \pm 0.007$ & $\textbf{0.6412} \pm 0.009$ & $\textbf{0.8537} \pm 0.007$\\ 
\hline
\end{tabular}
\label{tab:MSD-single}
\end{table*}

\noindent\textbf{Prostate158.} Prostate158 is a publicly available prostate MRI dataset that includes T2w, DWI, and ADC series collected from 158 patients. Radiologists have annotated the prostate regions and malignant tumor areas. We extract the prostate regions delineated by radiologists to serve as the final input images for the model.


\noindent\textbf{Experimental Details.} For all datasets, we randomly divide the whole set into training, validation, and testing as 8:1:1 patient-wise. While pretraining, we
randomly flip and crop a 96×96×96 volume as the input for BT-MSD, a 48×48×48 volume as the input for Breast2023 and Prostate158. Our pretraining employs 2 A6000 GPUs for 5 days, using Adam optimizer with a cosine decay learning rate starting at 1e-3. We run 8,000 epochs for BT-MSD, 2,000 epochs on Breast2023 and Prostate158. Transitioning to the fine-tuning phase, we adopt an initial learning rate of 5e-3, coupled with a weight decay of 0.01 to refine the model's performance across all tasks. And our training batch sizes are all 6. For each downstream task, we run 10 times and report average results and standard deviations.

\begin{figure*}[!t]
\includegraphics[width=\textwidth]{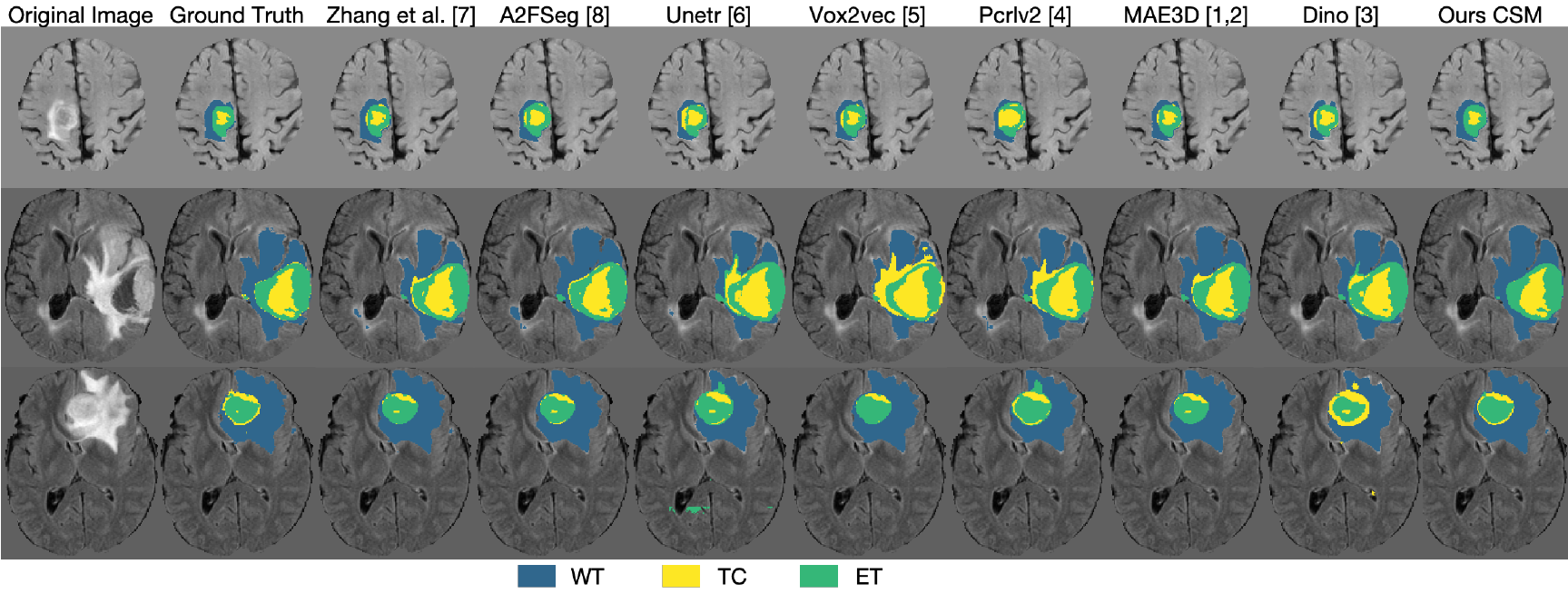}
\caption{The brain tumor segmentation results of all methods using multiple series.}
\label{brain_seg_result}
\end{figure*}

\begin{table*}[!t]
\caption{Breast MRI Malignant Diagnosis using single series and multiple series.}
\centering
\begin{tabular}{c|c|c}
\hline
 Methodology & AUC in single series& AUC in multiple series\\
\hline
ViT \cite{vit} & $0.7202 \pm 0.037$ & $0.7702 \pm 0.034$ \\
MAE3D \cite{zhouisbi,mae} & $0.7510 \pm 0.015$ & $0.7923 \pm 0.010$\\
Dino \cite{dino} & $0.7431 \pm 0.023$ & $0.7886 \pm 0.021$\\
Pcrlv2 \cite{pcrlv2} & $0.7455 \pm 0.017$ & $0.7817 \pm 0.019$\\
Vox2vec \cite{vox2vec} & $0.7498 \pm 0.018$ & $0.7869 \pm 0.015$\\ \cline{1-3}
\textbf{CSM (Ours)} & $\textbf{0.7887} \pm 0.006$ &  $\textbf{0.8407}  \pm 0.005$\\ \cline{1-3}
Zhang \textit{et al.} \cite{tmiattention} & - &  $0.7957 \pm 0.011$ \\
A2FSeg \cite{miccaiattention} & - & $0.7990 \pm 0.012$ \\\cline{1-3}
Zhang \textit{et al.} \cite{tmiattention} + \textbf{CSM (Ours)} & - & $\textbf{0.8545 } \pm 0.006$\\
A2FSeg \cite{miccaiattention} + \textbf{CSM (Ours)} & - &  $\textbf{0.8598} \pm 0.008$\\
\hline
\end{tabular}

\label{tab:Breast}
\end{table*}


\noindent\textbf{Compared Baselines.} We compare our model with several self-supervised learning methods: \textbf{a)} MAE3D \cite{zhouisbi,mae} utilize a masked autoencoder approach to improve feature extraction for medical imaging tasks. \textbf{b)} Dino \cite{dino} employs a teacher-student network paradigm to refine representation learning. \textbf{c)} Pcrlv2 \cite{pcrlv2} integrates pixel-level restoration to augment semantic features with finer details. \textbf{d)} Vox2vec \cite{vox2vec} adopts contrastive learning to enhance voxel-wise feature learning. The above self-supervised learning methods all focus on single-series pretraining without considering multi-series relationships. We also compare with several attention-based methods that consider multi-series fusion in supervised learning: \textbf{a)} Unetr \cite{unetre} employs a transformer-based encoder to capture global multi-scale information from 3D medical images for segmentation. \textbf{b)} ViT \cite{vit} uses a transformer-based encoder for classification. \textbf{c)} Zhang \textit{et al.} \cite{tmiattention} introduces a method that utilizes a flexible module to integrate features from multiple MRI series. \textbf{d)} A2FSeg \cite{miccaiattention} develops a feature fusion approach followed by a sophisticated adaptive fusion using an attention-based mechanism.

We use Adam to train our model. For all methods except Pcrlv2 \cite{pcrlv2} and Vox2vec \cite{vox2vec}, which are constrained by their original designs to use CNNs \cite{resnet}, the ViT-B/16 \cite{vit} is employed as the standard backbone.

\begin{table*}[!t]
\caption{Prostate Cancer Diagnosis using single series and multiple series.}
\centering
\begin{tabular}{c|c|c}
\hline
 Methodology & AUC in single series& AUC in multiple series\\
\hline
ViT \cite{vit} & $0.7013 \pm 0.031$ &  $0.7492 \pm 0.023$ \\
MAE3D \cite{zhouisbi,mae} &  $0.7316 \pm 0.021$ &  $0.8123 \pm 0.016$\\
Dino \cite{dino} & $0.7142 \pm 0.028$ &  $0.8077 \pm 0.020$\\
Pcrlv2 \cite{pcrlv2} &  $0.7239 \pm 0.022$ &  $0.8014 \pm 0.019$\\
Vox2vec \cite{vox2vec} &  $0.7230 \pm 0.025$ &  $0.8107 \pm 0.021$\\ \cline{1-3}
\textbf{CSM (Ours)} & $\textbf{0.7627} \pm 0.015$ &  $\textbf{0.8458}  \pm 0.013$\\ \cline{1-3}
Zhang \textit{et al.} \cite{tmiattention} & - &  $0.8179 \pm 0.018$\\
A2FSeg \cite{miccaiattention} & - & $0.8188 \pm 0.015$ \\\cline{1-3}
Zhang \textit{et al.} \cite{tmiattention} + \textbf{CSM (Ours)} & - & $\textbf{0.8492 } \pm 0.012$\\
A2FSeg \cite{miccaiattention} + \textbf{CSM (Ours)} & - &  $\textbf{0.8531} \pm 0.012$\\
\hline
\end{tabular}

\label{tab:Prostate}
\end{table*}

\begin{table*}[!t]
\caption{MSD Brain Tumor Segmentation under our proposed method CSM using multiple series by training with different amounts of annotated data.}
\centering
\begin{tabular}{c|c|c|c|c}
\hline
\multirow{2}{*}{Labeling Ratio} & \multicolumn{4}{c}{MSD Brain Tumor Segmentation} \\\cline{2-5}
 & Average & WT & ET & TC \\\hline
100\% & $0.8345 \pm 0.007$ & $0.9346 \pm 0.004$ & $0.6738 \pm 0.009$ & $0.8951 \pm 0.007$ \\
75\% & $0.8268 \pm 0.008$ & $0.9365 \pm 0.008$ & $0.6610 \pm 0.009$ & $0.8828 \pm 0.006$ \\
50\% & $0.7973 \pm 0.007$ & $0.9028 \pm 0.007$ & $0.6350 \pm 0.007$ & $0.8540 \pm 0.008$ \\
10\% & $0.7846 \pm 0.008$ & $0.8950 \pm 0.010$ & $0.6205 \pm 0.008$ & $0.8383 \pm 0.006$ \\

\hline
\end{tabular}

\label{tab:MSD-labelrate}
\end{table*}

\begin{table*}[!t]
\caption{Breast MRI Malignant Diagnosis and Prostate Cancer Diagnosis under our proposed method CSM using multiple series by training with different amounts of annotated data.}
\centering
\begin{tabular}{c|c|c}
\hline
Labeling Ratio & Breast MRI Malignant Diagnosis-AUC & Prostate Cancer Diagnosis-AUC \\\hline
100\% & $0.8407 \pm 0.005$ & $0.8458  \pm 0.013$\\
75\% & $0.8364 \pm 0.005$ & $0.8332 \pm 0.016$\\
50\% & $0.8182 \pm 0.008$ & $0.8251 \pm 0.018$\\
10\% & $0.7752 \pm 0.006$ & $0.8070 \pm 0.019$\\

\hline
\end{tabular}

\label{tab:BreastProstate-labelrate}
\end{table*}

\subsection{Results}

We verify our CSM on downstream multi-series tasks and single-series tasks. In our CSM, the encoder is trained in our Self-supervised Representation Learning stage with unlabeled multiple series. Then we use this well-trained encoder to initialize the encoder of each downstream task. Besides comparing with compared baselines, we also evaluate the effectiveness of combining CSM with the existing attention-based methods. The results are shown in Table~\ref{tab:MSD-multi}, Table~\ref{tab:MSD-single} Table~\ref{tab:Breast} and Table \ref{tab:Prostate}.

\noindent\textbf{BT-MSD.} From Table~\ref{tab:MSD-multi} and Table~\ref{tab:MSD-single}, our CSM outperforms the other methods on both multi-series and single-series MSD brain tumor segmentation, with an average Dice score of 0.8345 and 0.8013, respectively. This shows that our model can effectively capture the cross-series information and segment the brain tumor regions with higher accuracy. 

\noindent\textbf{Breast2023.} From Table~\ref{tab:Breast}, we can find that our model achieves the best performance on both single-series and multi-series breast MRI malignant diagnosis, with an AUC of 0.7887 and 0.8407, respectively. This demonstrates that our model can learn a robust representation from multiple series MRI data, and transfer it to a downstream task even if that is with a single series input. Note that the number of series in BT-MSD and Breast2023 is different (4 in BT-MSD and 3 in Breast2023), which demonstrates our ability to handle different numbers and types of series.

\noindent\textbf{Prostate158.} The results in Table \ref{tab:Prostate} illustrate the effectiveness of our CSM in prostate cancer diagnosis. Our CSM achieves an AUC of 0.7627 when analyzing single-series data and improves to 0.8458 with the integration of multi-series data. Our performances not only surpass other methods in both single-series and multi-series data but also highlight the benefits of integrating multiple MRI series for lesion diagnosis.

Furthermore, our model can also enhance the performance of Zhang \textit{et al.} \cite{tmiattention} and A2FSeg \cite{miccaiattention} by integrating our learned pre-trained model with their attention mechanisms in all tasks. The improved AUC scores suggest that our model can complement the existing attention-based methods by exploring more discriminative features. 

Additionally, our visualizations provide further evidence of the effectiveness of our CSM. Figure~\ref{brain_seg_result} shows the segmentation results of our CSM and other comparison methods in brain tumors. And Figure~\ref{cam} shows the class active maps of different series while learning multi-series representation in breast and prostate tumor benign/malignant classification under our CSM. These visualizations highlight the model's ability to focus on cross-series representation for robust diagnostic capabilities.

\noindent\textbf{Different Amounts of Annotated Data.} As shown in Table \ref{tab:MSD-labelrate} and Table \ref{tab:BreastProstate-labelrate}, we can see that our model can achieve competitive results on BT-MSD, Breast2023 and Prostate158 with different amounts of annotated data. This proves that our model can leverage the large-scale unlabeled MRI data to learn a generalizable representation and reduce the dependency on the labeled data. Even with only 10\% of the labeled data, our model can still achieve a decent performance on both tasks and have comparable performances with others using full datasets. This can demonstrate the potential of our model for practical applications with limited annotations.

\subsection{Ablation Study}

\begin{table*}[!t]
\caption{Ablation Study on MSD Brain Tumor Segmentation with our proposed method CSM using multiple series.}
\centering
\begin{tabular}{c|c|c|c|c|c|c|c}
\hline
 Mask-ratio & Random patch-mask & Series-mask & Complete-series & Average & WT & ET & TC  \\
\hline
87.5\% & $\checkmark$ & $\checkmark$ & $\checkmark$ & $0.8345 \pm 0.007$ & $0.9346 \pm 0.004$ & $0.6738 \pm 0.009$ & $0.8951 \pm 0.007$ \\
87.5\% & $\checkmark$ & $\circ$ & $\checkmark$ & $0.8309 \pm 0.007$ & $0.9296 \pm 0.007$ & $0.6736 \pm 0.009$ & $0.8894 \pm 0.005$\\
87.5\% & $\checkmark$ & $\times$ & $\checkmark$ & $0.8294 \pm 0.009$ & $0.9320 \pm 0.008$ & $0.6679 \pm 0.011$ & $0.8882 \pm 0.007$ \\
87.5\% & $\times$ & $\checkmark$ & $\checkmark$ & $0.7966 \pm 0.008$ & $0.9178 \pm 0.008$ & $0.6282 \pm 0.007$ & $0.8439 \pm 0.010$ \\
87.5\% & $\checkmark$ & $\checkmark$ & $\times$ & $0.8226 \pm 0.010$ & $0.9245 \pm 0.008$ & $0.6615 \pm 0.012$ & $0.8806 \pm 0.009$ \\
50\% & $\checkmark$ & $\checkmark$ & $\checkmark$ & $0.8325 \pm 0.007$ & $0.9239 \pm 0.006$ & $0.6906 \pm 0.008$ & $0.8820 \pm 0.007$ \\

\hline
\end{tabular}

\label{tab:MSD-ablation}
\end{table*}

\begin{table*}[!t]
\caption{Ablation Study on Breast MRI Malignant Diagnosis and Prostate Cancer Diagnosis with our proposed method CSM using multiple series.}
\centering
\begin{tabular}{c|c|c|c|c|c}
\hline
 Mask-ratio & Random patch-mask & Series-mask & Complete-series & Breast MRI Malignant Diagnosis-AUC & Prostate Cancer Diagnosis-AUC  \\
\hline
87.5\% & $\checkmark$ & $\checkmark$ & $\checkmark$ & $0.8407 \pm 0.005$ & $0.8458  \pm 0.013$\\
87.5\% & $\checkmark$ & $\circ$ & $\checkmark$ & $0.8350 \pm 0.005$ & $0.8420 \pm 0.012$\\
87.5\% & $\checkmark$ & $\times$ & $\checkmark$ & $0.8275 \pm 0.006$ & $0.8357 \pm 0.010$\\
87.5\% & $\times$ & $\checkmark$ & $\checkmark$ & $0.8195 \pm 0.007$ & $0.8196 \pm 0.016$\\
87.5\% & $\checkmark$ & $\checkmark$ & $\times$ & $0.8330 \pm 0.009$ & $0.8297 \pm 0.014$\\
50\% & $\checkmark$ & $\checkmark$ & $\checkmark$ & $0.8309 \pm 0.008$ & $0.8290 \pm 0.012$\\

\hline
\end{tabular}

\label{tab:BreastProstate-ablation}
\end{table*}

\noindent \textbf{Random patch-mask:} in \textit{Self-supervised Representation Learning} pretraining stage, whether randomly masking position of patches on different series. $\times$ represents masking patches in the same position on different series; $\checkmark$ represents masking patches in the random position on different series.

\noindent \textbf{Series-mask:} in \textit{Self-supervised Representation Learning} pretraining stage, whether randomly masking MRI series. $\times$ represents not masking any series, $\checkmark$ represents randomly masking partial series meanwhile reconstructing the masked series and $\circ$ represents randomly masking partial series but does not reconstruct the masked series.

\noindent \textbf{Complete-series:} in \textit{Downstream Finetuning} stage, $\checkmark$ represents all series in using multi-series data are complete. $\times$ represents that some series of some data in using multi-series data are missing (different subjects have different numbers of series). To ensure consistency in the test data, we artificially created such scenarios ($\times$) where some series of the data are missing by randomly eliminating one or two series. In finetuning and inference, such missing series can be padded with zeros to ensure uniformity of input length.

To evaluate the impact of different masking strategies and whether different subjects have different numbers of series on the performance using multiple series, we conduct the ablation study in Table~\ref{tab:MSD-ablation} for BT-MSD, Table~\ref{tab:BreastProstate-ablation} for Breast2023 and Prostate158. The study is structured around three main variables: the mask ratio, the masking strategy, and whether series are completely applied to the series or patches. We observed a high mask ratio of 87.5\% consistently yielded better performance compared to a lower mask ratio of 50\%. This suggests that a greater degree of information occlusion during training can lead to more robust cross-series representation learning.

The application of randomly patch-masking across different series (indicated by $\checkmark$) generally resulted in higher performance metrics than masking the same position of patches across series (indicated by $\times$). This indicates that introducing variability in the masked regions across series can be beneficial for the model's ability to generalize.
Furthermore, when comparing series-masking strategies, we find that randomly masking partial series while reconstructing the masked series (indicated by $\checkmark$) outperformed strategies where no series were masked (indicated by $\times$) and those where partial series were masked without reconstruction (indicated by $\circ$). This highlights the importance of reconstructing the masked series as it seems to encourage the model to learn more comprehensive and detailed representations.
The results of this ablation study provide valuable insights into the optimization of masking strategy for improving multi-series analysis.

In clinical practice, subjects typically may present with varying numbers of image series, a variability we have emulated in our study (indicated by $\times$). The comparative analysis in Table~\ref{tab:MSD-ablation} and Table~\ref{tab:BreastProstate-ablation} demonstrates that our method maintains efficacy with only a minor performance decrement between Row 5 and Row 1. Consequently, our approach holds promise for clinical applications.

\section{Conclusion} In this paper, we propose a novel method CSM for learning a robust representation from multiple series MRI data. CSM leverages cross-series masking to encourage the model to learn common and complementary information from different MRI series. We evaluate our method and achieve state-of-the-art performances on three downstream tasks: MSD brain tumor segmentation, breast MRI cancer and prostate cancer diagnosis. Moreover, CSM can achieve competitive results with different amounts of annotated data, demonstrating its potential for practical applications with limited annotations.


\begin{thebibliography}{00}

\bibitem{zhouisbi}
L. Zhou, H. Liu, J. Bae, J. He, D. Samaras, and P. Prasanna, "Self Pre-Training with Masked Autoencoders for Medical Image Classification and Segmentation," in \textit{Proc. IEEE 20th Int. Symp. Biomed. Imaging (ISBI)}, Cartagena, Colombia, pp. 1-6, 2023, doi: 10.1109/ISBI53787.2023.10230477.

\bibitem{mae}
K. He, X. Chen, S. Xie, Y. Li, P. Dollár, and R. Girshick, "Masked Autoencoders are Scalable Vision Learners," in \textit{Proc. IEEE/CVF Conf. Comput. Vis. Pattern Recognit.}, pp. 16000-16009, 2022.

\bibitem{dino}
M. Caron, H. Touvron, I. Misra, H. Jégou, J. Mairal, P. Bojanowski, and A. Joulin, "Emerging Properties in Self-supervised Vision Transformers," in \textit{Proc.IEEE/CVF Int.Conf. Comput. Vis.}, pp. 9650-9660, 2021.

\bibitem{pcrlv2}
H.-Y. Zhou, C. Lu, C. Chen, S. Yang, and Y. Yu, "A Unified Visual Information Preservation Framework for Self-supervised Pre-Training in Medical Image Analysis," in \textit{IEEE Trans. Pattern Anal. Mach. Intell.}, vol. 45, no. 7, pp. 8020-8035, Jul. 1, 2023, doi: 10.1109/TPAMI.2023.3234002.

\bibitem{vox2vec}
M. Goncharov, V. Soboleva, A. Kurmukov, M. Pisov, and M. Belyaev, "Vox2vec: A Framework for Self-supervised Contrastive Learning of Voxel-level Representations in Medical Images," in \textit{Proc. Int. Conf. Med. Image Comput. Comput.-Assist. Interv.}, pp. 605-614, Cham, Springer Nature Switzerland, Oct. 2023.

\bibitem{unetre}
A. Hatamizadeh, Y. Tang, V. Nath, D. Yang, A. Myronenko, B. Landman, H. R. Roth, and D. Xu, "Unetre: Transformers for 3D Medical Image Segmentation," in \textit{Proc. IEEE/CVF Winter Conf. Appl. Comput. Vis.}, pp. 574-584, 2022.

\bibitem{tmiattention}
Y. Zhang, C. Peng, R. Tong, L. Lin, Y. W. Chen, Q. Chen, H. Hu, and SK. Zhou, "Multi-Modal Tumor Segmentation With Deformable Aggregation and Uncertain Region Inpainting," in \textit{IEEE Trans. Med. Imaging}, vol. 42, no. 10, pp. 3091-3103, Oct. 2023, doi: 10.1109/TMI.2023.3275592.

\bibitem{miccaiattention}
Z. Wang and Y. Hong, "A2FSeg: Adaptive Multi-modal Fusion Network for Medical Image Segmentation," in \textit{Proc. Int. Conf. Med. Image Comput. Comput.-Assist. Interv.}, pp. 673-681, Cham, Springer Nature Switzerland, Oct. 2023.

\bibitem{vit}
A. Dosovitskiy, L. Beyer, A. Kolesnikov, D. Weissenborn, X. Zhai, T. Unterthiner, M. Dehghani, M. Minderer, G. Heigold, S. Gelly, J. Uszkoreit, and N. Houlsby, "An Image is Worth 16x16 Words: Transformers for Image Recognition at Scale," in \textit{Proc. Int. Conf. Learn. Represent.}, 2020, pp. 1-18.

\bibitem{old1}
J. J. Corso, E. Sharon, S. Dube, S. El-Saden, U. Sinha, and A. Yuille, "Efficient Multilevel Brain Tumor Segmentation With Integrated Bayesian Model Classification," \textit{IEEE Trans. Med. Imaging}, vol. 27, no. 5, pp. 629-640, May 2008, doi: 10.1109/TMI.2007.912817.

\bibitem{old2}
J. Zhou, K. L. Chan, V. F. H. Chong, and S. M. Krishnan, "Extraction of Brain Tumor from MR Images Using One-Class Support Vector Machine," in \textit{Proc. Annu. Int. Conf. IEEE Eng. Med. Biol. Soc.}, Shanghai, China, pp. 6411-6414, 2005, doi: 10.1109/IEMBS.2005.1615965.

\bibitem{mriback1}
S. Lee, S. Y. Lee, J. Y. Jung, Y. Nam, H. J. Jeon, C. K. Jung, S. H. Shin, and Y. Chung, "Ensemble Learning-based Radiomics with Multi-sequence Magnetic Resonance Imaging for Benign and Malignant Soft Tissue Tumor Differentiation," in \textit{PLoS One}, vol.18, e0286417, 2023.

\bibitem{mriback2}
S. Boudabbous, M. Hamard, E. Saiji, K. Gorican, P. A. Poletti, M. Becker, and A. Neroladaki, "What Morphological MRI Features Enable Differentiation of Low-Grade from High-Grade Soft Tissue Sarcoma?," in \textit{BJR| Open}, vol. 4, no. 1, 20210081, 2022.

\bibitem{mriback3}
H. Wang, H. Chen, S. Duan, D. Hao, and J. Liu, "Radiomics and Machine Learning With Multiparametric Preoperative MRI May Accurately Predict the Histopathological Grades of Soft Tissue Sarcomas," in \textit{J. Magn. Reson. Imaging}, vol. 51, no. 3, pp. 791-797, Mar. 2020.

\bibitem{mriback4}
M. J. Emaus, M. F. Bakker, P. H. Peeters, C. E. Loo, R. M. Mann, M. D. De Jong, R. H. Bisschops, J. Veltman, K. M. Duvivier, M. B. Lobbes, R. M. Pijnappel, N. Karssemeijer, H. J. Koning, M. A. Bosch, E. M. Monninkhof, W. P. Mali, W. B. Veldhuis, and C. H. Gils, "MR Imaging as an Additional Screening Modality for the Detection of Breast Cancer in Women Aged 50–75 Years with Extremely Dense Breasts: The DENSE Trial Study Design," in \textit{Radiology}, vol.277, pp. 527-537, 2015.

\bibitem{mriback5}
E. A. Sickles. "ACR BI-RADS® Atlas, Breast imaging reporting and data system," American College of Radiology, 2013: 39.

\bibitem{old3}
H. Cai, R. Verma, Y. Ou, S.-K. Lee, E. R. Melhem, and C. Davatzikos, "Probabilistic Segmentation of Brain Tumors Based on Multi-Modality Magnetic Resonance Images," in \textit{Proc. 4th IEEE Int. Symp. Biomed. Imaging: From Nano to Macro}, Arlington, VA, USA, pp. 600-603, 2007, doi: 10.1109/ISBI.2007.356923.

\bibitem{brats}
M. Antonelli, A. Reinke, S. Bakas, K. Farahani, A. Kopp-Schneider, B. A. Landman, G. Litjens, B. Menze, O. Ronneberger, R. M. Summers, B. V. Ginneken, M. Bilello, P. Bilic, P. F. Christ, R. K. Do, M. J. Gollub, S. H. Heckers, H. Huisman, W. R. Jarnagin, M. K. McHugo, S. Napel, J. S. Pernicka, K. Rhode, C. Tobon-Gomez, E. Vorontsov, J. A. Meakin, S. Ourselin, M. Wiesenfarth, P. Arbeláez, B. Bae, S. Chen, L. Daza, J. Feng, B. He, F. Isensee, Y. Ji, F. Jia, I. Kim, K. Maier-Hein, D. Merhof, A. Pai, B. Park, M. Perslev, R. Rezaiifar, O. Rippel, I. Sarasua, W. Shen, J. Son, C. Wachinger, L. Wang, Y. Wang, Y. Xia, D. Xu, Z. Xu, Y. Zheng, A. L. Simpson, L. Maier-Hein, and M. J. Cardoso, "The Medical Segmentation Decathlon," in \textit{Nat. Commun.}, vol. 13, no. 1, 4128, 2022, doi: 10.1038/s41467-022-29979-8.

\bibitem{resnet}
K. He, X. Zhang, S. Ren, and J. Sun, "Deep Residual Learning for Image Recognition," in \textit{Proc. IEEE Conf. Comput. Vis. Pattern Recognit.}, pp. 770-778, 2016.

\bibitem{bookmri}
M. T. Vlaardingerbroek and J. A. Boer, "Magnetic Resonance Imaging: Theory and Practice," in \textit{Science \& Business Media}, Springer, 2013.

\bibitem{prostate_Roussel}
E. Roussel, U. Capitanio, A. Kutikov, E. Oosterwijk, I. Pedrosa, S. P. Rowe, and M. A. Gorin, "Novel Imaging Methods for Renal Mass Characterization: A Collaborative Review," in \textit{Eur. Urol.}, vol. 81, no. 5, pp. 476-488, 2022.

\bibitem{prostate_Wysock}
J. S. Wysock and H. Lepor, "Multi-Parametric MRI Imaging of the Prostate—Implications for Focal Therapy," in \textit{Transl. Androl. Urol.}, vol. 6, no. 3, pp. 453, 2017.

\bibitem{bladder_Panebianco}
V. Panebianco, Y. Narumi, E. Altun, B. H. Bochner, J. A. Efstathiou, S. Hafeez, R. Huddart, S. Kennish, S. Lerner, R. Montironi, V. F. Muglia, G. Salomon, S. Thomas, H. A. Vargas, J. A. Witjes, M. Takeuchi, J. Barentsz, and J. W. Catto, "Multiparametric Magnetic Resonance Imaging for Bladder Cancer: Development of VI-RADS (Vesical Imaging-Reporting And Data System)," in \textit{Eur. Urol.}, vol. 74, no. 3, pp. 294-306, 2018.

\bibitem{glioblastoma_Shukla}
G. Shukla, G. S. Alexander, S. Bakas, R. Nikam, K. Talekar, J. D. Palmer, and W. Shi, "Advanced Magnetic Resonance Imaging in Glioblastoma: A Review," in \textit{Chin. Clin. Oncol.},vol.6, pp.40-40,2017.

\bibitem{global_cancer2022}
F. Bray, M. Laversanne, H. Sung, J. Ferlay, R. L. Siegel, I. Soerjomataram, A. Jemal, "Global Cancer Statistics 2022: GLOBOCAN Estimates of Incidence and Mortality Worldwide for 36 Cancers in 185 Countries," in \textit{CA: A Cancer Journal for Clinicians}, vol. 74, no. 3, pp. 229-263, May-Jun. 2024.

\bibitem{mpmri_Winfield}
J. M. Winfield, G. S. Payne, A. Weller, and N. M. deSouza, "DCE-MRI, DW-MRI, and MRS in Cancer: Challenges and Advantages of Implementing Qualitative and Quantitative Multi-Parametric Imaging in the Clinic," in \textit{Top. Magn. Reson. Imaging}, vol.25, pp.245-254, 2016.

\bibitem{mpmri_Omari}
E. A. Omari, Y. Zhang, E. Ahunbay, E. Paulson, A. Amjad, X. Chen, Y. Liang, X. A. Li, "Multi‐Parametric Magnetic Resonance Imaging for Radiation Treatment Planning," in \textit{Med. Phys.}, vol. 49, no. 4, pp. 2836-2845, 2022.

\bibitem{breast_mann}
R. M. Mann, N. Cho, and L. Moy, "Breast MRI: State of the Art," in \textit{Radiology}, vol. 292, no. 3, pp. 520-536, Sep. 2019.

\bibitem{breast_pinker}
K. Pinker, T. H. Helbich, and E. A. Morris, "The Potential of Multiparametric MRI of the Breast," in \textit{Br. J. Radiol.}, vol. 90, no. 1069, 20160715, 2017.

\bibitem{simple_chen}
T. Chen, S. Kornblith, M. Norouzi, and G. Hinton, "A Simple Framework for Contrastive Learning of Visual Representations," in \textit{Proc. Int. Conf. Mach. Learn.} (ICML), pp. 1597-1607, 2020.

\bibitem{big_chen}
T. Chen, S. Kornblith, K. Swersky, M. Norouzi, and G. E. Hinton, "Big Self-Supervised Models are Strong Semi-Supervised Learners," in \textit{Adv. Neural Inf. Process. Syst.} (NeurIPS), vol. 33, pp. 22243-22255, 2020.

\bibitem{empirical_chen}
X. Chen, S. Xie, and K. He, "An empirical study of training self-supervised vision transformers," in \textit{Proc. IEEE/CVF Int. Conf. Comput. Vis.} (ICCV), pp. 9640-9649, 2021.

\bibitem{bootstrap_grill}
J. B. Grill, F. Strub, F. Altché, C. Tallec, P. Richemond, E. Buchatskaya, C. Doersch, B. A. Pires, Z. Guo, M. G. Azar, B. Piot, K. kavukcuoglu, R. Munos, and M. Valko, "Bootstrap Your Own Latent: A New Approach to Self-Supervised Learning," in \textit{Adv. Neural Inf. Process. Syst.}, vol. 33, pp. 21271-21284, 2020.

\bibitem{momentum_he}
K. He, H. Fan, Y. Wu, S. Xie, and R. Girshick, "Momentum Contrast for Unsupervised Visual Representation Learning," in \textit{Proc. IEEE/CVF Conf. Comput. Vis. Pattern Recognit.} (CVPR), pp. 9729-9738, 2020.

\bibitem{siamese_chen}
X. Chen and K. He, "Exploring Simple Siamese Representation Learning," in \textit{Proc. IEEE/CVF Conf. Comput. Vis. Pattern Recognit.} (CVPR), pp. 15750-15758, 2021.

\bibitem{unsupervised_caron}
M. Caron, I. Misra, J. Mairal, P. Goyal, P. Bojanowski, and A. Joulin, "Unsupervised Learning of Visual Features by Contrasting Cluster Assignments," in \textit{Adv. Neural Inf. Process. Syst.} (NeurIPS), vol. 33, pp. 9912-9924, 2020.

\bibitem{selfsupervised_wen}
X. Wen, B. Zhao, A. Zheng, X. Zhang, and X. Qi, "Self-Supervised Visual Representation Learning with Semantic Grouping," in \textit{Adv. Neural Inf. Process. Syst.} (NeurIPS), vol. 35, pp. 16423-16438, 2022.

\bibitem{simple_xie}
Z. Xie, Z. Zhang, Y. Cao, Y. Lin, J. Bao, Z. Yao, Q. Dai, and H. Hu, "SimMIM: A Simple Framework for Masked Image Modeling," in \textit{Proc. IEEE/CVF Conf. Comput. Vis. Pattern Recognit.} (CVPR), pp. 9653-9663, 2022.

\bibitem{hard_wang}
H. Wang, K. Song, J. Fan, Y. Wang, J. Xie, and Z. Zhang, "Hard patches mining for masked image modeling," in \textit{Proc. IEEE/CVF Conf. Comput. Vis. Pattern Recognit.} (CVPR), pp. 10375-10385, 2023.

\bibitem{peco_dong}
X. Dong, J. Bao, T. Zhang, D. Chen, W. Zhang, L. Yuan, D. Chen, F. Wen, N. Yu, and B. Guo, "Peco: Perceptual Codebook for BERT Pre-training of Vision Transformers," in \textit{Proc. AAAI Conf. Artif. Intell.} (AAAI), vol. 37, no. 1, pp. 552-560, Jun. 2023.

\bibitem{neural_oord}
A. van den Oord and O. Vinyals, "Neural discrete representation learning," in \textit{Adv. Neural Inf. Process. Syst.} (NeurIPS), vol. 30, 2017.

\bibitem{masked_wei}
C. Wei, H. Fan, S. Xie, C. Y. Wu, A. Yuille, and C. Feichtenhofer, "Masked Feature Prediction for Self-supervised Visual Pre-training," in \textit{Proc. IEEE/CVF Conf. Comput. Vis. Pattern Recognit.} (CVPR), pp. 14668-14678, 2022.

\bibitem{semmae_li}
G. Li, H. Zheng, D. Liu, C. Wang, B. Su, and C. Zheng, "Semmae: Semantic-guided Masking for Learning Masked Autoencoders," in \textit{Adv. Neural Inf. Process. Syst.} (NeurIPS), vol. 35, pp. 14290-14302, 2022.

\bibitem{adversarial_shi}
Y. Shi, N. Siddharth, P. Torr, and A. R. Kosiorek, "Adversarial Masking for Self-supervised Learning," in \textit{Proc. Int. Conf. Mach. Learn.} (ICML), pp. 20026-20040, Jun. 2022, PMLR.

\bibitem{mriback6_prostate}
B. Turkbey, A. B. Rosenkrantz, M. A. Haider, A. R. Padhani, G. Villeirs, K. J. Macura, C. M. Tempany, P. L. Choyke, F. Cornud, D. J. Margolis, and H. C. Thoeny, "Prostate Imaging Reporting and Data System Version 2.1: 2019 Update of Prostate Imaging Reporting and Data System Version2," in \textit{Eur. Urol.}, vol.76, no.3, pp.340-351, 2019.

\bibitem{mixedae}
K. Chen, Z. Liu, L. Hong, H. Xu, Z. Li, and D. Y. Yeung, "Mixed Autoencoder for Self-Supervised Visual Representation Learning," in \textit{Proc. IEEE/CVF Conf. Comput. Vis. Pattern Recognit.} (CVPR), pp. 22742-22751, 2023.

\bibitem{modelgenesis}
Z. Zhou, V. Sodha, J. Pang, M. B. Gotway, and J. Liang, "Models Genesis," in \textit{Med. Image Anal.}, vol. 67, 101840, 2021.

\bibitem{rubikscube}
J. Zhu, Y. Li, Y. Hu, K. Ma, S. K. Zhou, and Y. Zheng, "Rubik's Cube+: A Self-Supervised Feature Learning Framework for 3D Medical Image Analysis," in \textit{Med. Image Anal.}, vol. 64, 101746, 2020.

\bibitem{swinunetr}
Y. Tang, D. Yang, W. Li, H. R. Roth, B. Landman, D. Xu, V. Nath, A. Hatamizadeh, "Self-Supervised Pre-Training of Swin Transformers for 3D Medical Image Analysis," in \textit{Proc. IEEE/CVF Conf. Comput. Vis. Pattern Recognit.} (CVPR), pp. 20730-20740, 2022.



\bibitem{xie2020pgl}
Y. Xie, J. Zhang, Z. Liao, Y. Xia, and C. Shen, "PGL: Prior-Guided Local Self-Supervised Learning for 3D Medical Image Segmentation," in \textit{arXiv preprint}, arXiv:2011.12640, 2020.

\bibitem{ye2022desd}
Y. Ye, J. Zhang, Z. Chen, and Y. Xia, "DeSD: Self-Supervised Learning with Deep Self-Distillation for 3D Medical Image Segmentation," in \textit{Proc. Int. Conf. Med. Image Comput. Comput.-Assist. Interv.}, pp. 545-555, Cham, Springer Nature Switzerland, Sep. 2022.

\bibitem{zhou2021preservational}
H. Y. Zhou, C. Lu, S. Yang, X. Han, and Y. Yu, "Preservational Learning Improves Self-Supervised Medical Image Models by Reconstructing Diverse Contexts," in \textit{Proc. IEEE/CVF Int. Conf. Comput. Vis.} (ICCV), pp. 3499-3509, 2021.

\bibitem{wang2024fremim}
W. Wang, J. Wang, C. Chen, J. Jiao, Y. Cai, S. Song, and J. Li, "FreMIM: Fourier Transform Meets Masked Image Modeling for Medical Image Segmentation," in \textit{Proc. IEEE/CVF Winter Conf. Appl. Comput. Vis.} (WACV), pp. 7860-7870, 2024.



\end{thebibliography}
\end{document}